\documentclass[10pt,conference]{IEEEtran}
\IEEEoverridecommandlockouts

\usepackage[utf8]{inputenc}
\usepackage{cite}
\usepackage{amsmath,amssymb,amsfonts}
\usepackage{algorithmic,algorithm}
\usepackage{graphicx}
\usepackage{textcomp,xcolor}
\usepackage{booktabs,multirow,tabularx}
\usepackage{tcolorbox}
\usepackage{xspace,enumitem}
\usepackage{xurl}
\usepackage[hidelinks]{hyperref}

\newcommand{\this}{\textbf{KnotScryer}\xspace}

\newcommand{\etal}{\textit{et al.}}
\definecolor{g}{HTML}{009F4D}
\definecolor{b}{HTML}{2529D8}
\definecolor{r}{HTML}{DE0F17}
\definecolor{darkred}{HTML}{8B0000}
\definecolor{darkgreen}{HTML}{007800}

\hypersetup{
  pdftitle={Detecting Insecure Compositions of Clean Agent Skills via Capability-Flow Graph Analysis},
  pdfauthor={Yihao Li, Chuanpu Fu, Yizhou Zhao, Miao Chen, Qi Li, Ke Xu},
  pdfsubject={KnotScryer: security analysis of cross-skill compositions in AI agents},
  pdfkeywords={agent security, cross-skill composition, capability-flow analysis, KnotScryer}
}

\title{Detecting Insecure Compositions of Clean Agent Skills via Capability-Flow Graph Analysis}
\author{%
  \IEEEauthorblockN{Yihao Li\textsuperscript{1}\quad
  Chuanpu Fu\textsuperscript{2}\quad
  Yizhou Zhao\textsuperscript{3}\quad
  Miao Chen\textsuperscript{4}\quad
  Qi Li\textsuperscript{1}\quad
  Ke Xu\textsuperscript{1,*}}
  \IEEEauthorblockA{\small\textsuperscript{1}Tsinghua University\\
  \small\textsuperscript{2}Nanyang Technological University\\
  \small\textsuperscript{3}Institute for Network Sciences and Cyberspace, Tsinghua University\\
  \small\textsuperscript{4}Zhongguancun Laboratory}
  \thanks{\textsuperscript{*}Corresponding author: Ke Xu.}%
}

\date{}

\begin{document}
\maketitle
\begin{abstract}

AI agents increasingly extend their capabilities by invoking third-party skills, where each skill encapsulates instructions, scripts, and resources for agents to execute on demand. 
This paper reveals that deploying many benign skills together introduces compositional security risks: multiple individually clean skills can compose into safety-violation chains, because the agent passes an upstream skill's output to a downstream skill's input across invocations. 
The risk arises not from any single skill, but from the agent's runtime orchestration, and thus cannot be captured by existing malicious skill detection methods. 

To this end, we develop \this to detect insecure skill compositions by treating skills as typed capability flows. 
In particular, we abstract each skill into a capability summary, and construct the Skill Capability-Flow Graph (SCFG), where edges denote cross-skill information flow. 
Moreover, we design a search policy that tracks protected or untrusted data along handoff edges to locate paths reaching insecure operations.
Finally, a two-layer verification method replays the insecure information flow for reachability analysis and applies a task-anchored LLM chain audit to confirm the semantic plausibility of the identified violation chains.

We deploy \this to analyze 13{,}598 clean skills from ClawHub, and validate the detected chains across three agent frameworks and three LLMs.
\this identifies 612 insecure compositions, which are consistently exploited across agent frameworks. Meanwhile, under GLM-5.1, it outperforms tag-match and graph-only baselines by improving the exploitation rate from 16.5\% $\sim$ 18.2\% to 85.6\%, while reducing the search space by 85.3\%.
Furthermore, 13.7\% of the detected chains consist entirely of skills labeled safe by a single-skill scanner, showing that \this effectively identifies insecure compositions that per-skill auditing cannot.

\end{abstract}

\section{Introduction}
\label{sec:introduction}

Large language models (LLMs) are evolving from passive conversation into AI agents capable of invoking tools to execute complex tasks.
To extend AI agents' task capabilities, agent ecosystems have introduced the \emph{skill} mechanism~\cite{anthropic-skills}.
A skill encapsulates natural-language instructions, scripts, and data, packaged into a reusable software unit for agents to install, load, and execute on demand.
Since its introduction in October 2025, this mechanism has been standardized and rapidly  adopted. It has been released as an open Agent Skills standard~\cite{anthropic-skills-standard} and integrated into agent platforms such as Claude Code and Cursor~\cite{thenewstack-skills, aibusiness-skills}.
As of June~2026, SkillsMP, a general-purpose skill registry, indexes over 1.9 million public skill files crawled from GitHub~\cite{skillsmp}, while ClawHub, a curated marketplace with security review~\cite{agensi-marketplaces}, hosts over 65K skills.

However, directly invoking third-party skills exposes agents to security risks~\cite{SP26-AgenticSoK,NDSS26-ToolSelection,CCS25-AgentSentinel}.
The existing mainstream skill platforms have deployed security inspection before a skill is published.
For example, ClawHub runs static scans over each submitted skill---inspecting its scripts and declared permissions---and labels only those passing all checks as \emph{clean}~\cite{agensi-marketplaces}.
Similarly, open scanners such as Cisco AI Defense's Skill Scanner combine pattern matching and LLM-assisted review to detect malicious code and prompt in individual skills~\cite{Cisco-SkillScanner}.
In addition, recent works further detect malicious skills in the wild, revealing skills that conceal harmful behaviors from users~\cite{SEC26-MaliciousSkill}.
While these efforts prevent the propagation of \emph{individually} malicious skills from a single artifact, we instead ask a different question: analogous to return-oriented programming (ROP)~\cite{CCS07-ROP,SEC25-GadgetChain,SEC25-PcodeCRA}, which assembles benign code gadgets into a malicious one, can attackers compose \emph{benign} skills into a chain that threatens the agent's security?

This paper discloses insecure compositions of \emph{clean} skills.
Even when every skill is labeled clean with no malicious logic under marketplace audit, they can still form executable safety violations under the agent's runtime orchestration.
Because the agent automatically forwards one skill's output into another skill's input across invocations, therefore the composed data flow may reach a forbidden operation, such as unauthorized code execution.
For instance, consider an agent with two \emph{clean} ClawHub skills installed: (i) \texttt{gitlab-cli-skills}, which inspects GitLab repositories and issues; and (ii) \texttt{deploy-on-render}, which automatically installs and executes a project.
We observe that, when the agent triggers the first skill to inspect a GitLab project, explicit instructions in public issues can consistently trigger the second skill that installs and executes a designated project, allowing attackers to obtain code execution privilege on the agent's host.
In general, the security threat stems not from any individual skill, but from the agent's runtime composition of clean skills.

\begin{figure}[htbp]
\centering
\includegraphics[width=\columnwidth]{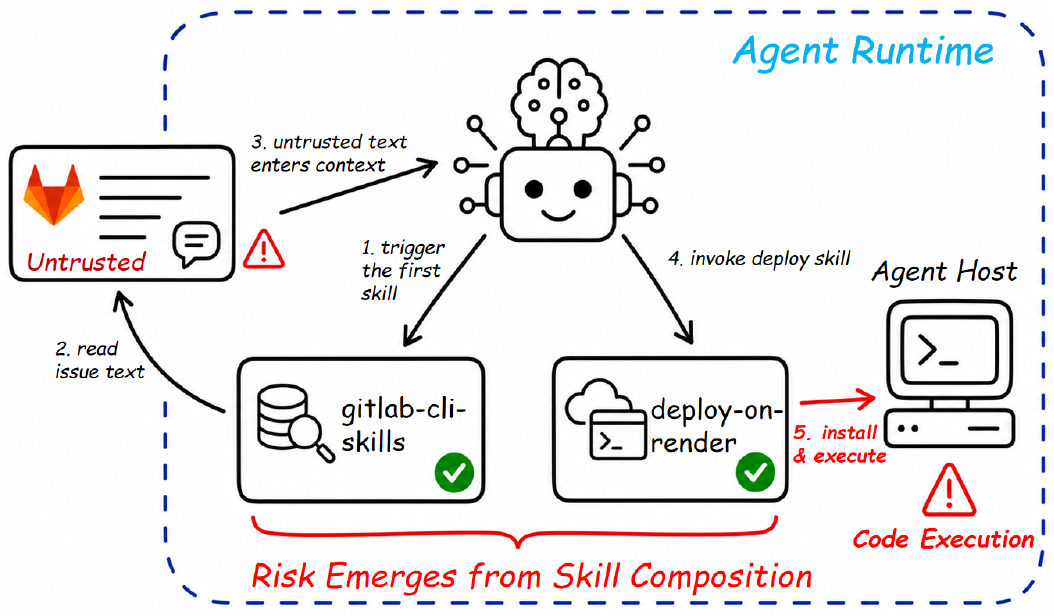}
\caption{Motivating example. Two marketplace-clean skills compose into a code execution chain through the agent's runtime orchestration.}
\label{fig:motivating}
\end{figure}

To systematically discover insecure skill compositions that can be exploited by attackers, this paper proposes \this, which constructs a graph to analyze the potential insecure transition among skills, and efficiently searches for dangerous compositions from the nearly unbounded space of skill combinations.
In particular, \this first summarizes the capability of each skill, and then constructs the \emph{Skill Capability-Flow Graph (SCFG)} based on these capabilities, where nodes denote skills and edges denote potential information-flow relationships between them.
Afterward, we design an efficient graph search method that automatically identifies information- and control-flow compositions that may violate security specifications.
Finally, we further filter candidate chains through two-layer verification: marker-flow reachability verification and task-anchored LLM chain audit.

To assess the effectiveness of \this, we apply it to 13{,}598 clean skills crawled via the ClawHub API (i.e., skills that have passed the platform's security review) and validate the detected insecure composition across three mainstream agent frameworks and three LLMs.
The experimental results show that \this detects 612 high-confidence violation chains, with an average exploitation rate of 79.2\% across the nine agent-LLM combinations.
In particular, under GLM-5.1, \this achieves an 85.6\% exploitation rate, while the tag-match and graph-only baselines reach only 16.5\% and 18.2\%, respectively.
Moreover, the exploited chains are consistent rather than runtime-specific: 45.5\%~$\sim$~46.3\% of them are reproduced across all three frameworks or all three models.
Furthermore, we validate that all the chains cannot be identified by existing commercial per-skill auditing tools.

In summary, the contributions of this paper are three-fold:
\begin{itemize}
    \item \textbf{New Problem.} We disclose insecure compositions of \emph{clean} agent skills, where individually benign, audited skills are composed by the agent at runtime into executable violations that skill auditing cannot capture.
    \item \textbf{New Algorithm.} We propose \this, which utilizes a capability-flow graph to search the vast composition space for dangerous skill chains, and further filters candidate chains through two-layer verification.
    \item \textbf{New Observation.} On 13{,}598 clean ClawHub skills, we uncover 612 violation chains that are 79.2\% exploitable across nine agent-LLM combinations, which are invisible to skill scanning tools.
\end{itemize}

\section{Background and Problem Setting}
\label{sec:background}

\subsection{Skill Composition}
\label{subsec:skill-composition}

A \emph{skill} is the basic unit for extending AI agents' capabilities. A skill typically includes three components:

\begin{enumerate}[nosep, leftmargin=*]
    \item \textit{SKILL.md.}~A natural-language description of the skill's functionality, trigger conditions, input/output formats, and usage constraints. The agent reads SKILL.md to decide whether and when to invoke the skill.
    \item \textit{Executable scripts.}~Code that implements the skill's concrete functionality (e.g., Python scripts, shell scripts), invoked by the agent at runtime.
    \item \textit{Templates and resources.}~The skill may include prompt templates, configuration files, or data resources for parameterizing the skill's behavior.
\end{enumerate}

Unlike traditional software packages (e.g., npm packages, PyPI libraries), a skill uniquely combines \emph{natural-language semantics} and \emph{executable code}. 
The description in SKILL.md is read and reasoned about not only by human developers but also by the agent's LLM directly, forming the semantic basis for skill selection and orchestration. 
To prevent the propagation of skills containing malicious logic (either in text or code form), some existing public skill platforms have begun to provide skill scanning services. 
For example, ClawHub labels skills as \textit{Clean} (a.k.a. \textit{Pass}), highlight \textit{Suspicious} skills, and reject malicious skills, which is similar to other commercial tools, e.g., Cisco SkillScanner~\cite{Cisco-SkillScanner}.

\subsection{Agent Skill Workflow and Agent-Mediated Handoff}
\label{subsec:handoff}

When an agent processes a user task, each skill invocation follows a natural cycle: the agent decides which skill in a unified list to call, the skill executes, and the agent reasons about the output. We call this the \textbf{DER sequence} (Decision $\to$ Execution $\to$ Reasoning).

\begin{enumerate}[nosep, leftmargin=*, labelindent=0pt, align=left]
    \item \textbf{Decision.}~The agent selects which skills to invoke based on the task and the skills' semantic descriptions. This selection is dynamic and context-dependent.
    \item \textbf{Execution.}~The agent executes the selected skill. Each skill's execution is independently authorized: the agent verifies per-skill invocation permissions but does not check cross-skill composition permissions.
    \item \textbf{Reasoning.}~The agent processes the skill's output, integrates it into its context, and determines subsequent actions based on the observed content.
\end{enumerate}

\begin{figure}[htbp]
\centering
\includegraphics[width=\columnwidth]{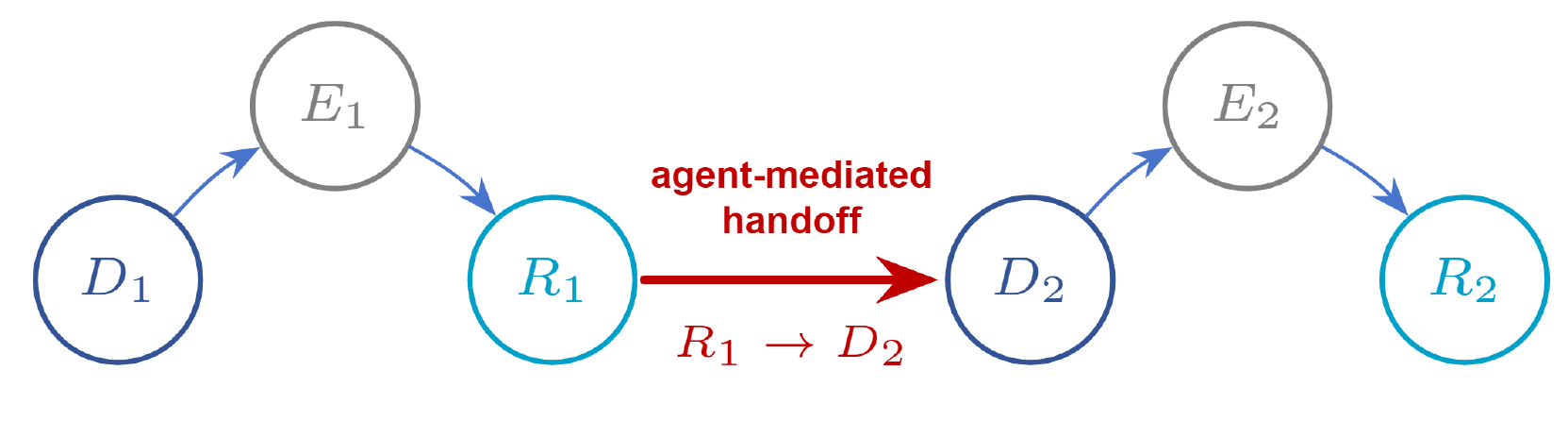}
\caption{The DER sequence and agent-mediated handoff. One DER  (D$\to$E$\to$R) represents a single skill invocation; the handoff ($R_1 \to D_2$) represents the agent-mediated data transfer between consecutive invocations.}
\label{fig:rec-model}
\end{figure}

Between two invocations, the agent's reasoning function determines the next skill to call, based on the previous skill's output. 
This process is defined as \textit{agent-mediated handoff}. As shown in Figure~\ref{fig:rec-model}, after reasoning about the upstream output ($R_1$), the agent decides to invoke a downstream skill ($D_2$) and repeats the execution-reasoning-decision loop for the invocation of the next skill.
An invocation chain is a sequence of skills connected by agent-mediated handoffs ($D_1 \wedge E_1 \wedge R_1 \wedge D_2 \wedge E_2 \wedge \cdots$). In general, the execution results of upstream skills can influence the handoff that determines which downstream skill in the chain will be executed.

\subsection{Problem Statement}
\label{subsec:problem}
We observe that, through agent-mediated handoffs, the output of an upstream skill can influence the agent's decision on whether and how a downstream skill is invoked. 
Since this output may carry malformed content, a low-privilege upstream skill (e.g., a retrieval-and-analysis skill) can be exploited to induce the agent to invoke a downstream high-privilege skill (e.g., one that executes a program and uploads information), forming an abnormal invocation chain.
This study aims to find insecure skill compositions that may constitute such violation chains.

The goal of \this is to discover skill composition that an agent can construct at runtime into a \emph{violation chain}.
We require that every skill is itself free of malicious logic, i.e., no skill is flagged as malicious by the existing skill scanning tools~\cite{Cisco-SkillScanner,agensi-marketplaces}.

We further require that every violation chain output by \this possess structural executability, i.e., the upstream skill can consistently drive the invocation of the downstream skill, rather than a skill that the agent may never sequentially execute at runtime.

Moreover, \this should be efficient enough to analyze the massive number of skills hosted on public skill platforms such as ClawHub.
Finally, the discovered insecure compositions are actionable for defense: their constituent skills can be prevented from being simultaneously visible in the agent's skill list during execution, or users can be advised against installing such unsafe composition together.

\subsection{Threat Model}
\label{subsec:threat-model}

We consider an \emph{external} attacker who cannot manipulate user inputs, cannot directly publish malicious skills, and cannot modify the agent framework. 
The attacker can only observe the skills available on public platforms, especially the popular ones with high download counts. 
Based on such observation, the attacker hosts static resources and plants crafted content within them, waiting to be retrieved and analyzed by the public skills. 
For example, the attacker posts comments under trending topics on social platforms (e.g., to be pulled by \texttt{x-twitter-scraper}), submits issues or merge requests on code platforms (e.g., to be inspected by \texttt{gitlab-cli-skills}), or hosts static web pages (e.g., to be fetched by  \texttt{rss-fetch} skill), attempting to be retrieved during skill execution.
In this way, the attacker interferes with the user agent's selection of skills within an unsafe composition. 
The injected content is then handed off step by step across multiple skills---from a low-privilege skill that first retrieves it, through intermediate skills that relay it to a high-privilege skill that consumes it. 
Ultimately, it reaches a high-privilege operation and triggers abnormal execution logic, such as data exfiltration, code execution, or prompt injection.

\begin{figure}[htbp]
\centering
\includegraphics[width=\columnwidth]{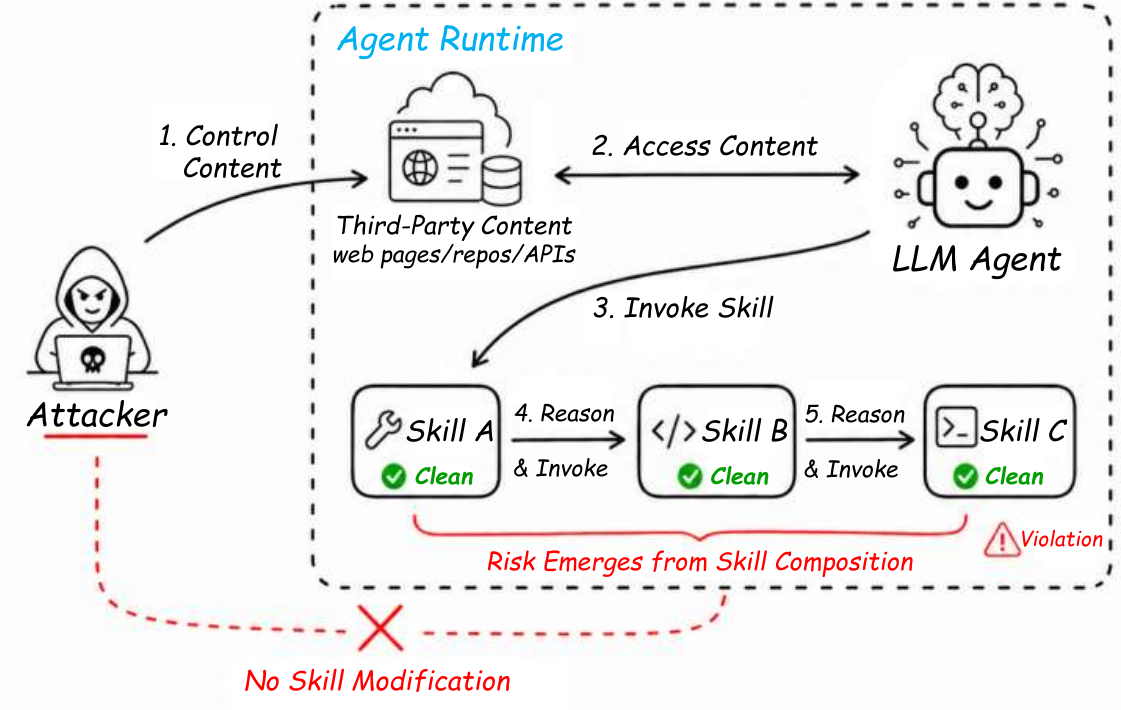}
\caption{Threat model. The attacker does not modify any skill. The risk emerges from the agent-mediated composition of independently benign skills.}
\label{fig:threat-model}
\end{figure}

\section{Approach}
\label{sec:approach}

\subsection{Overview}
\label{subsec:overview}

KnotScryer is built around one core design judgment: type compatibility is a necessary condition for composition, not a sufficient one. Only when the upstream output is observable to the agent, the downstream input is controllable by the agent, and both ends have type and evidence support, does an analyzable \textbf{agent-mediated handoff} exist between two skills.

This judgment formalizes the handoff conditions identified in Section~\ref{subsec:handoff} into a static analysis: rather than predicting whether the agent will reason about an output or decide to invoke a skill, KnotScryer checks whether the observable, controllable, and type-compatible conditions for a handoff are in place, with evidence supporting each claim.

The overall pipeline proceeds through four stages sequentially. SCS extracts the receiving, producing, and affecting capabilities of each individual skill. SCFG determines whether agent-mediated cross-skill handoffs exist among these capabilities. Policy search locates subsets of handoff paths that reach policy-forbidden operations. Verification confirms whether candidate paths have complete data channels and whether the entire handoff chain can be strung together by a coherent task.

\begin{figure*}[t]
\centering
\includegraphics[width=\textwidth]{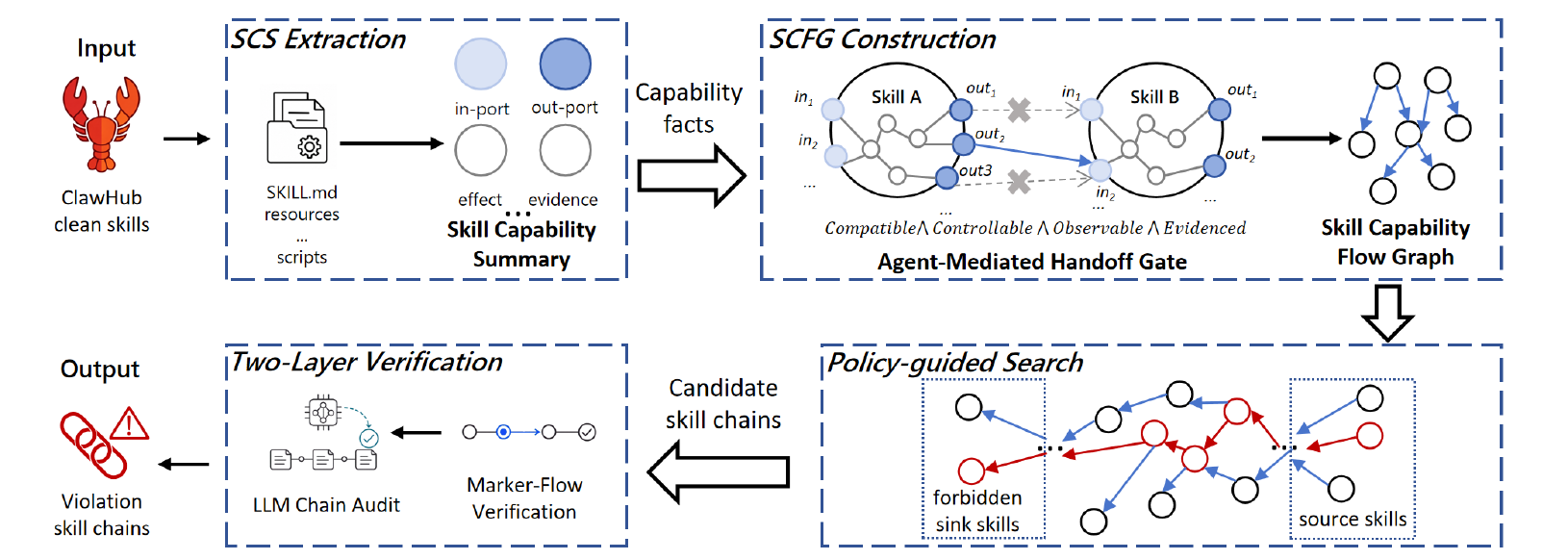}
\caption{Overview of KnotScryer.}
\label{fig:pipeline}
\end{figure*}

\subsection{Skill Capability Summary (SCS)}
\label{subsec:scs}

Agent skills are not ordinary software packages with only function signatures. A skill may simultaneously contain natural-language instructions, scripts, templates, resource files, and operations on the agent context. Retaining only names, descriptions, or risk tags loses the inputs, outputs, visibility, and side effects needed for cross-skill reasoning. We therefore design the \emph{Skill Capability Summary} (SCS) to model each skill by what it can receive, produce, and affect. SCS combines YAML frontmatter parsing, natural-language heuristic matching, and script static scanning, covering structured declarations, unstructured descriptions, code-level behaviors, and semantic properties that resist rule-based capture, respectively.

For each skill, SCS extracts six categories of information. \textsc{Input/output ports} record data type, sensitivity, trust origin, exposure scope, and consumability per port. Only outputs exposed to the agent context or workspace can participate in agent-mediated handoff. \textsc{Effects} record security-related side effects produced when the skill processes data, carrying flags for sensitivity preservation, trust-boundary crossing, and authorization requirements. Authorization requirements are recorded separately from input controllability and jointly determine whether a path reaches an unauthorized sink during  search and verification. \textsc{Operations} record the main actions the skill exposes and their parameters, supporting controllability assessment. \textsc{Context exposure} records visibility scopes including the agent context, workspace, external, and persistent state, supporting observability judgment. \textsc{Trust boundaries} record cross-domain relationships among local, third-party, internet, and persistent state, supporting trust-domain tracking. \textsc{Evidence} records source file, snippet, extraction method, and confidence score per claim.

\subsection{Skill Capability-Flow Graph (SCFG)}
\label{subsec:scfg}

Based on SCS, we construct a port-level Skill Capability-Flow Graph. Let the graph be $G = (V, E)$. Nodes include skills, input ports, output ports, effects, operations, context nodes, and trust-boundary nodes. Intra-skill edges describe reads, produces, transforms, exposes, sanitizes, and crosses-boundary relationships. The defining feature of SCFG is the \textbf{agent-mediated handoff conditions for cross-skill edges}: SCFG connects only those skill pairs that are simultaneously type-compatible, observable, controllable, and evidenced, i.e., runtime-reachable handoff paths.

Let $o$ be an output port of skill $s_i$ and $i$ be an input port of skill $s_j$. SCFG establishes an agent-mediated handoff edge only when all of the following conditions hold simultaneously:
\begin{equation}
\label{eq:handoff}
\begin{aligned}
(o, i) \in E_{\mathit{handoff}} \iff {} & \mathit{Compatible}(o, i)\;\wedge \\
  & \mathit{Observable}(o)\;\wedge\; \mathit{Controllable}(i)\;\wedge \\
  & \mathit{Evidenced}(o, i)
\end{aligned}
\end{equation}

Type compatibility ($\mathit{Compatible}$): the upstream output's data type can be consumed by the downstream input. Identical types connect directly; cross-type relationships are declared by an explicit compatibility table, e.g., URL can be consumed by Text, rather than defaulting to treating all outputs as interchangeable text.

Upstream observability ($\mathit{Observable}$): the output is presented to the agent as a tool return value, standard output, summary, transcript, workspace artifact, or explicit context exposure. Data existing only in the skill's private state (LocalOnly) cannot form agent-mediated handoff.

Downstream controllability ($\mathit{Controllable}$): the agent can provide task text, prompt content, command-line arguments, API request bodies, or workspace content as the downstream input. Controllability measures whether the agent can supply the input; whether the operation requires explicit authorization is recorded as a separate effect attribute and determines whether the path constitutes an unauthorized forbidden sink during policy search.

Evidence sufficiency ($\mathit{Evidenced}$): observability and controllability must each be supported by SCS evidence from the upstream and downstream skills, respectively. Descriptive guesses or empty evidence cannot independently establish high-confidence edges.

Note that this equivalence is about edge membership, not about whether a violation exists; a handoff edge only means that the data channel is in place, and whether it constitutes a violation still depends on policy search (Section~\ref{subsec:policy-search}).

This definition makes SCFG model the \emph{static necessary conditions} for runtime handoff: it checks not only type compatibility but also requires the agent to actually be in a position for handoff, i.e., observable and controllable, with supporting evidence, rather than assuming by default that the agent can copy arbitrary content.

\noindent\textbf{Scalable edge construction.}~Graph construction proceeds in two steps: first adding each skill's internal nodes and edges, then establishing cross-skill handoff edges under the four conditions above. To control graph density on large corpora, the cross-skill step indexes input ports by data type and checks only port pairs within compatible type buckets; it then executes exposure, controllability, and evidence checks, and imposes a per-output-port upper bound $b$ on cross-skill edges, prioritizing downstream ports that contain policy-relevant dangerous effects.

\begin{algorithm}[htbp]
\caption{SCFG Construction}
\label{alg:scfg}
\begin{algorithmic}[1]
\REQUIRE Set of SCS summaries $\mathcal{S}$, type compatibility table $\mathcal{T}$, per-output bound $b$
\ENSURE Capability-flow graph $G = (V, E)$
\STATE \textbf{// Step 1: Add intra-skill structure}
\STATE $V \leftarrow \emptyset$, $E \leftarrow \emptyset$
\FOR{each skill $s \in \mathcal{S}$}
  \STATE Add $s$, internal nodes, and intra-skill edges to $V$, $E$
\ENDFOR
\STATE \textbf{// Step 2: Add cross-skill handoff edges (type-indexed)}
\STATE $E_{\mathit{handoff}} \leftarrow \emptyset$
\STATE $\mathit{Idx}[t] \leftarrow \{i \mid i \text{ is an input port of type } t\}$
\FOR{each output port $o$ across all skills}
  \STATE $C_o \leftarrow \bigcup_{t' \in \mathit{CompatibleTypes}(\mathit{type}(o), \mathcal{T})} \mathit{Idx}[t']$
  \STATE $H_o \leftarrow \{(o, i) \mid i \in C_o \text{ from a different skill } \wedge\; \mathit{Observable}(o) \wedge \mathit{Controllable}(i) \wedge \mathit{Evidenced}(o, i)\}$
  \STATE Add top-$b$ of $H_o$ sorted by policy-relevant effect priority to $E_{\mathit{handoff}}$
\ENDFOR
\STATE $E \leftarrow E \cup E_{\mathit{handoff}}$
\RETURN $G = (V, E)$
\end{algorithmic}
\end{algorithm}

As Algorithm~\ref{alg:scfg} shows, Step~2 checks all four handoff conditions (Eq.~\eqref{eq:handoff}): type compatibility is pre-filtered by the type index into per-bucket traversal, and the remaining three conditions ($\mathit{Observable}$, $\mathit{Controllable}$, $\mathit{Evidenced}$) are explicitly checked on candidate port pairs. Together they ensure that only paths where the agent can plausibly hand off data at runtime, with supporting evidence, enter the graph. 

\subsection{Policy-Driven Capability-Flow Search}
\label{subsec:policy-search}

However, not every handoff path in SCFG constitutes a risk. A document-processing skill passing text to a summarization skill is a normal composition; a path becomes a candidate violation chain only when protected data or untrusted content reaches a policy-forbidden effect along the path. We therefore express security requirements as graph-search policies.

\subsubsection{Policy Definition}
\label{subsubsec:policy-def}

Each policy defines three things: what source conditions to start tracking from, what operations constitute forbidden sinks, and what sanitization or authorization conditions can terminate propagation. We instantiate four policies covering common risk patterns (Table~\ref{tab:policies}).

\begin{table*}[t]
\centering
\caption{Four security policy instances. Each specifies source, sink, and permitted blocking conditions.}
\label{tab:policies}
\small
\begin{tabularx}{\linewidth}{@{}lXXX@{}}
\toprule
\textbf{Policy} & \textbf{Source} & \textbf{Forbidden Sink} & \textbf{Blocking Condition} \\
\midrule
Data Exfiltration (DE) & Private, Secret, DerivedSensitive data & Unauthorized network send or external sink & Sanitization, declassification, or explicit authorization \\
Indirect Prompt Injection (IPI) & Internet, ThirdParty untrusted content & Agent context or task execution & Content sanitization or context isolation \\
Code Execution Propagation (CEP) & Code or Command data & Shell or agent task execution & Explicit authorization \\
Fact Poisoning (FP) & Content with manipulative semantics & Memory, configuration, or persistent-state write & Sanitization or write authorization \\
\bottomrule
\end{tabularx}
\end{table*}

These four policies cover common risk patterns: DE covers sensitive data propagation and trust boundary crossing, IPI covers third-party content elevation, CEP covers narrow capability accumulation into high impact execution, and FP covers persistent state write risks. Adding a new policy requires only declaring new trigger conditions and state transitions without changing the graph structure or search algorithm.

\subsubsection{Policy State Tracking and Search-Space Control}
\label{subsubsec:search-process}

Given an SCFG and a set of policies, the search aims to find paths that start from protected data or untrusted content and reach policy-forbidden sinks along handoff edges. Consider the data exfiltration policy (DE): the search starts from output ports marked Private, tracks data propagation along handoff edges; when the path passes through sanitization or explicit authorization, the safe branch terminates; when the path reaches an unauthorized network-send effect, the search reports a candidate violation chain.

We formalize this process as follows. Let the capability-flow graph be $G = (V, E)$ and the policy be $P = (Q, \delta, q_0, F)$:
\begin{itemize}\setlength{\itemsep}{1pt}
\item $Q = \{\mathit{Initial}, \mathit{Tracking}, \mathit{Propagation}, \mathit{Violation}, \mathit{Safe}\}$ --- the set of policy states;
\item $F = \{\mathit{Violation}\}$ --- the violation states;
\item $q_0 = \mathit{Initial}$ --- the initial state;
\item $\delta$ --- the state transition function.
\end{itemize}
\noindent The search does not explicitly construct the full product graph; instead, it maintains policy state synchronously during traversal of $G$. Let $\Delta \subseteq (V \times Q) \times (V \times Q)$ denote the transition relation of the implicitly maintained product of $G$ and $P$. A transition satisfies:
\begin{equation}
\label{eq:product}
((v, q),\, (v', q')) \in \Delta \iff (v, v') \in E \;\wedge\; q' = \delta(q, \sigma(v, v')),
\end{equation}
where $\sigma(v, v')$ represents the sensitivity level, trust origin, exposure scope, effect type, and authorization information carried by the node or edge. The function $\delta$ advances state according to the security attributes observed along the path. Under DE, for example: when the path first passes a Private port, the policy moves from \emph{Initial} to \emph{Tracking}; when the Private data reaches a controllable downstream input, it moves to \emph{Propagation}; when the propagation reaches an unauthorized network-send effect, it moves to \emph{Violation}. At any state, if the path encounters sanitization, declassification, explicit authorization, or LocalOnly blocking, $\delta$ transitions to \emph{Safe} and the search stops expanding that branch.

The search starts from output ports that satisfy the policy's source condition and whose evidence meets the confidence threshold. Each search state consists of three components: the current graph node, the policy state, and \emph{provenance information} recording which skill and port the data originally came from, along with the sensitivity level and trust domain at that point. This provenance-aware design prevents erroneously merging paths from different skills: even if two paths reach the same node, as long as their origins differ, they may correspond to different responsibility subjects and mitigation locations.

\begin{algorithm}[htbp]
\caption{Policy-Guided Search with Provenance Tracking}
\label{alg:search}
\begin{algorithmic}[1]
\REQUIRE SCFG $G$, policies $\{P_j\}_{j=1}^{r}$, per-policy budget $K_j$, max depth $D$
\ENSURE Set of candidate violation chains $\mathcal{C}$
\STATE $\mathcal{C} \leftarrow \emptyset$
\FOR{each policy $P_j$}
  \STATE $\mathit{Worklist} \leftarrow \{(v, q_{0,j}, \mathit{prov}_0, 0) \mid v \text{ satisfies source condition}\}$
  \WHILE{$\mathit{Worklist} \neq \emptyset$ \AND $|\mathcal{C}_j| < K_j$}
    \STATE Pop $(v, q, \mathit{prov}, d)$ from worklist
    \IF{$d > D$} \STATE \textbf{continue} \ENDIF
    \IF{$q \in F_j$} \STATE Add chain with provenance $\mathit{prov}$ to $\mathcal{C}_j$; \textbf{continue} \ENDIF
    \FOR{each $(v, v') \in E$}
      \STATE $q' \leftarrow \delta_j(q, \sigma(v, v'))$ \hfill \textcolor{darkred}{\textbf{// policy-driven transition}}
      \IF{$q'$ is safe} \STATE \textbf{continue} \ENDIF
      \IF{$(v', q', \mathit{prov})$ dominated by existing state} \STATE \textbf{continue} \ENDIF \hfill \textcolor{darkred}{\textbf{// provenance-aware pruning}}
      \STATE Push $(v', q', \mathit{prov} \cup \{(v, q, \sigma)\}, d{+}1)$ to worklist
    \ENDFOR
  \ENDWHILE
  \STATE $\mathcal{C} \leftarrow \mathcal{C} \cup \mathcal{C}_j$
\ENDFOR
\RETURN $\mathcal{C}$
\end{algorithmic}
\end{algorithm}

The search space is controlled by four mechanisms: type and handoff checks delete impossible edges at graph-construction time, reducing the number of candidate pairs; sanitization and authorization states terminate safe branches early, avoiding continued expansion on already-mitigated paths; provenance-aware dominance pruning merges states that share both origin and policy state; maximum depth and per-policy budget $K_j$ prevent any single policy from expanding indefinitely.

\subsubsection{Candidate Chain Output and Complexity}
\label{subsubsec:bounded-search}

Candidate chains retain the information needed for verification: the skill sequence and port path for replaying the data flow, the policy state trajectory and risk reason for determining the violation type, and the provenance evidence for pinpointing responsibility. Under DE, for example, a candidate chain records which skill's output port the Private data originates from, which handoffs it passes through, and which unauthorized network-send effect it reaches, enabling the subsequent marker-flow verification to inject markers along the same path and check reachability.

If policy $j$ retains $S_j$ non-dominated states and expands $A_j$ transitions in practice, the actual search time for $r$ policies is
\begin{equation}
\label{eq:search-complexity}
O\!\left(\sum_{j=1}^{r} (S_j + A_j)\right).
\end{equation}
Under the approximation of ignoring provenance splitting, this becomes $O(r |Q| (|V| + |E|))$. The depth and budget limits guarantee that the search terminates in a finite number of steps.

\subsection{Two-Layer Verification}
\label{subsec:verification}

Policy search finds violation paths on the graph, but two types of uncertainty remain: (1)~a candidate may satisfy the policy only at local nodes, with data not actually reaching the forbidden sink along the full path; (2)~adjacent handoffs may each be individually valid, but the entire chain may not be stringable together by a single plausible task. KnotScryer addresses these with deterministic marker-flow verification and task-anchored LLM chain audit, respectively.

\subsubsection{Marker-Flow Reachability Verification}
\label{subsubsec:marker-flow}

For each candidate chain, the validator injects a synthetic marker (e.g., untrusted-content) at the source and replays propagation along the saved capability-flow trajectory. A marker must survive until it reaches a concrete effect; if blocked by sanitization, declassification, or an authorization gate before the sink, the path does not constitute an executable violation. Marker flow thus provides deterministic, reproducible reachability evidence.

\subsubsection{Task-Anchored LLM Chain Audit}
\label{subsubsec:llm-audit}

Whereas SCFG checks structural properties, the LLM-auditor judges at the semantic level whether the chain is executable: whether the output is semantically accessible to the agent, whether the controllable input would be triggered by a plausible task, and whether adjacent handoffs jointly form the same task. The auditor uses a fixed-prompt LLM and can only confirm or downgrade SCFG-discovered chains, without modifying the graph or adding candidates.

\section{Evaluation}
\label{sec:evaluation}

In this section, we evaluate SCFG's ability to identify exploitable violation chains, remain consistent across agents and LLMs, outperform baseline methods, and scale with acceptable overhead. We also conduct ablation studies on the handoff conditions and present an end-to-end case study.

\subsection{Experimental Setup}
\label{subsec:setup}

\textbf{Dataset Construction.}~We crawl all skills labeled \emph{clean} in the ClawHub 2026-05-18 snapshot, totaling 13{,}598 skills. Discovering cross-skill violations on a reviewed platform is the most stringent test: if the platform has weaker review, the risk only increases. We also extract a \emph{popular} subgraph (Stars $\geq$ 20 or Downloads $\geq$ 1{,}500), yielding 1{,}062 skills.

\textbf{Agent frameworks and LLMs.}~We select three open-source agent frameworks: Hermes (v0.16.0), OpenClaw (2026.6.8), and OpenCode (v1.17.7), and test each with three LLMs: Claude Sonnet 4.6, GPT-5.5, and GLM-5.1. The frameworks differ in skill selection strategy, execution authorization, and context processing; the LLMs differ in reasoning capability and safety alignment. The $3 \times 3$ cross tests generalizability across both dimensions.

\textbf{Exploitation strategy.}~Our strategy aligns with the threat model (Section~\ref{subsec:threat-model}): the attacker controls third-party content sources accessed by upstream skills but cannot directly control agent execution. For each detected violation chain, we inject attacker-influenced data at the source skill's third-party content source, provide the agent with a generic task instruction that merely triggers the source skill, and observe whether the agent, upon receiving the source skill's influenced output through normal reasoning, autonomously propagates along the violation chain to execute the sink skill---i.e., whether the $R_1 \!\to\! D_2$ handoff (Section~\ref{subsec:handoff}) occurs without direct attacker control over the agent's execution. The case study in Section~\ref{subsec:rq4} complements the batch experiments by demonstrating a complete end-to-end attack scenario with a natural user prompt. Each chain is executed at most twice per agent; if either attempt succeeds, the chain is marked as genuinely present.

\textbf{Ethical considerations.}~For safety concerns, we do not expose any manipulated content that can trigger the violation skill chains to the Internet. Instead, all experiments are conducted in a fully isolated local environment: one machine simulates the attacker, hosting the third-party content sources and data-receiving endpoints, and another simulates the user, running the agent with the installed skills. All traffic stays within the local network, and no real users, systems, or third-party services are involved. We have responsibly disclosed the discovered violation chains to the skill platform.

\subsection{Detection Results and Exploitability}
\label{subsec:rq1}

SCFG detects 612 high-confidence violation chains on the full 13{,}598-skill corpus. We execute attack experiments on all 612 chains; Table~\ref{tab:exploitation} reports the results.

\begin{table}[htbp]
\centering
\caption{Exploitation results and consistency analysis for 612 SCFG-detected violation chains.}
\label{tab:exploitation}
\small
\begin{tabularx}{\columnwidth}{@{}l*{4}{>{\centering\arraybackslash}X}@{}}
\toprule
& \multicolumn{4}{c}{Exploitation Rate (\%)} \\
\cmidrule(lr){2-5}
& GLM-5.1 & Sonnet 4.6 & GPT-5.5 & Overall \\
\midrule
Hermes   & 92.2 & 65.0 & 82.8 & 80.0 \\
OpenClaw & 71.7 & 92.3 & 51.1 & 71.7 \\
OpenCode & 93.0 & 68.0 & 96.9 & 86.0 \\
\midrule
\textbf{Average} & \textbf{85.6} & \textbf{75.1} & \textbf{77.0} & \textbf{79.2} \\
\midrule
& \multicolumn{4}{c}{Cross-Agent Consistency (\%)} \\
\cmidrule(lr){2-5}
& All 3 & Exactly 2 & Exactly 1 & None \\
\midrule
GLM-5.1    & 62.6 & 32.2 & 4.7 & 0.5 \\
Sonnet 4.6 & 35.6 & 54.7 & 9.0 & 0.7 \\
GPT-5.5    & 38.2 & 57.2 & 1.8 & 2.8 \\
\midrule
\textbf{Average} & \textbf{45.5} & \textbf{48.0} & \textbf{5.2} & \textbf{1.3} \\
\midrule
& \multicolumn{4}{c}{Cross-LLM Consistency (\%)} \\
\cmidrule(lr){2-5}
& All 3 & Exactly 2 & Exactly 1 & None \\
\midrule
Hermes   & 46.7 & 46.6 & 6.7 & 0.0 \\
OpenClaw & 31.5 & 52.6 & 15.4 & 0.5 \\
OpenCode & 60.8 & 36.3 & 2.9 & 0.0 \\
\midrule
\textbf{Average} & \textbf{46.3} & \textbf{45.2} & \textbf{8.3} & \textbf{0.2} \\
\bottomrule
\end{tabularx}
\end{table}

\textbf{Overall exploitation rate.}~Across $3 \times 3 = 9$ agent-LLM combinations, the 612 SCFG-detected chains show high exploitability. The three LLM backends achieve overall rates of GLM-5.1 (85.6\%), Sonnet 4.6 (75.1\%), and GPT-5.5 (77.0\%), with a weighted overall of 79.2\%. For cross-agent consistency, the All-3 rates are 62.6\% (GLM-5.1), 35.6\% (Sonnet 4.6), and 38.2\% (GPT-5.5); zero-exploitation chains account for only 0.5\%, 0.7\%, and 2.8\%, respectively. Cross-LLM consistency is likewise significant: 60.8\% of chains are exploited by all three models on OpenCode, 46.7\% on Hermes, and 31.5\% on OpenClaw. These results confirm that SCFG-detected violations represent cross-framework, cross-model structural risks rather than runtime-specific artifacts.

\textbf{Framework differences.}~The three frameworks show different overall exploitation rates: OpenCode (86.0\%) $>$ Hermes (80.0\%) $>$ OpenClaw (71.7\%), with cross-LLM consistency following the same order. Combining framework implementation details with execution logs, we find that OpenClaw mediates skill invocations through a gateway in relevant invocation paths and provides SSRF protection for certain network-request-class operations, while Hermes and OpenCode do not employ such framework-level mediation. This difference in defense depth causes security outcomes to vary across frameworks. The results indicate that framework-level protection can change the threshold and failure modes of risk, but is insufficient to independently eliminate cross-skill composition risks.

\textbf{Model differences.}~The three LLM backends rank as GLM-5.1 (85.6\%) $>$ GPT-5.5 (77.0\%) $>$ Sonnet 4.6 (75.1\%). Cross-agent consistency reveals deeper divergence: GLM-5.1 achieves the highest All-3 rate at 62.6\%, while Sonnet 4.6 and GPT-5.5 show 35.6\% and 38.2\%, respectively. Refusals by Sonnet 4.6 and GPT-5.5 stem from model safety alignment proactively aborting execution upon recognizing danger signals in context (e.g., external endpoint requests, injection instructions). However, safety alignment takes effect only when the payload is directly exposed and struggles to block attacks transmitted indirectly through skill chains. Model-side defense alone is insufficient to eliminate cross-skill composition risks.

\subsection{Comparison with Baselines}
\label{subsec:rq2}

We select two baselines that represent alternative approaches to cross-skill risk analysis. \emph{Tag-match} is inspired by SkillProbe~\cite{skillprobe2026}: it assigns coarse risk tags to skills and connects source risk tags with sink/input tags, capturing skill-level associations. \emph{Graph-only} is inspired by Graph of Skills~\cite{graphskills2026}: it builds a typed-port dependency graph and checks reachability via type compatibility and graph structure.

Both baselines are built on the popular subgraph (1{,}062 skills), as the combination space on the full corpus would be intractable. Even on this small corpus, the combination space explodes: tag-match produces 829,462 associations and graph-only produces 1,047,207 edges. The root cause is that neither baseline checks whether data can actually flow between two skills via the agent. Tag-match links any skill carrying a source risk tag with any skill carrying a sink tag, regardless of whether the output is visible to the agent or the input is agent-suppliable. Graph-only connects all type-compatible ports and propagates along reachable paths, but type compatibility alone does not imply agent observability or controllability. Both methods thus treat composition as a pairwise tag or type relation, not as an agent-mediated channel. SCFG, in contrast, requires all four handoff conditions (Eq.~\eqref{eq:handoff}) to hold simultaneously: on the same subgraph, it produces only 13 candidate chains, constraining the search space to agent-executable semantics. On the popular subgraph, all 13 SCFG candidate chains are exploited under GLM-5.1 on at least one agent framework.

For exploitability evaluation, exhaustively testing the large number of baseline candidates is infeasible. We therefore sample 500 candidates from each baseline uniformly at random and test each against three agents (using GLM-5.1) under the same attack strategy, reporting 95\% confidence intervals (CI) for the exploitation rate. Statistical significance is assessed via two-proportion $z$-test with continuity correction.

\begin{table}[htbp]
\centering
\caption{Exploitation results: SCFG vs.\ baselines, broken down by agents.}
\label{tab:baseline}
\small
\setlength{\tabcolsep}{2.5pt}
\begin{tabularx}{\columnwidth}{@{}llrrr>{\centering\arraybackslash}X@{}}
\toprule
Method & Agent & Exploited & Total & Rate & 95\% CI \\
\midrule
\multirow{4}{*}{Tag-match} & Hermes & 204 & 500 & 40.8\% & [36.5\%,\,45.1\%] \\
 & OpenClaw & 35 & 500 & 7.0\% & [\phantom{0}4.8\%,\,\phantom{0}9.2\%] \\
 & OpenCode & 8 & 500 & 1.6\% & [\phantom{0}0.5\%,\,\phantom{0}2.7\%] \\
\cmidrule(lr){2-6}
 & \textbf{Overall} & \textbf{247} & \textbf{1{,}500} & \textbf{16.5\%} & \textbf{[14.7\%,\,18.4\%]} \\
\midrule
\multirow{4}{*}{Graph-only} & Hermes & 220 & 500 & 44.0\% & [39.6\%,\,48.4\%] \\
 & OpenClaw & 47 & 500 & 9.4\% & [\phantom{0}6.8\%,\,12.0\%] \\
 & OpenCode & 6 & 500 & 1.2\% & [\phantom{0}0.2\%,\,\phantom{0}2.2\%] \\
\cmidrule(lr){2-6}
 & \textbf{Overall} & \textbf{273} & \textbf{1{,}500} & \textbf{18.2\%} & \textbf{[16.3\%,\,20.2\%]} \\
\midrule
\multirow{4}{*}{SCFG} & Hermes & 564 & 612 & 92.2\% & \multicolumn{1}{c}{---} \\
 & OpenClaw & 439 & 612 & 71.7\% & \multicolumn{1}{c}{---} \\
 & OpenCode & 569 & 612 & 93.0\% & \multicolumn{1}{c}{---} \\
\cmidrule(lr){2-6}
 & \textbf{Overall} & \textbf{1{,}572} & \textbf{1{,}836} & \textbf{85.6\%} & \multicolumn{1}{c}{---} \\
\bottomrule
\end{tabularx}
\end{table}

Both baselines show low exploitation rates (Table~\ref{tab:baseline}): tag-match 16.5\% and graph-only 18.2\%. The 95\% confidence intervals ([14.7\%,\,18.4\%] and [16.3\%,\,20.2\%]) confirm that even in the best case, baseline exploitation remains well below SCFG's 85.6\%. The difference is statistically significant ($z > 37$, $p < 0.001$; i.e., the gap is not due to sampling variance). These results validate the earlier hypothesis: tag and type matching connect skills without checking agent-mediated channel conditions, so the vast majority of their candidates are not agent-executable.

\textbf{Cross-agent consistency.}~Under GLM-5.1, SCFG's cross-agent consistency far exceeds the baselines. SCFG achieves a 62.6\% All-3 consistency rate, while both baselines show extremely low All-3 rates. Hermes exploits baseline candidates (tag-match 40.8\%, graph-only 44.0\%), but OpenClaw and OpenCode barely exploit any (1.6\%--9.4\%), meaning most baseline candidates are exploited by at most one framework or none at all. This indicates that baseline candidates lack the structural guarantees of agent-mediated handoff, making their exploitability vary with framework configuration. SCFG candidates, by satisfying handoff conditions, exhibit structural consistency across frameworks.

To further demonstrate the necessity of cross-skill analysis, we align the 612 LLM-confirmed SCFG chains with per-skill labels from a third-party Cisco AI Defense single-skill scanner. Among them, 84 chains (13.7\%) have all skills labeled safe by Cisco, meaning a portion of composition risk is completely invisible to component-level scanning. The remaining 522 chains contain at least one skill labeled unsafe. These chains reflect a granularity mismatch. Component-level alerts can only flag locally suspicious skills. They cannot reconstruct source-to-sink cross-skill violation chains. An additional 6 chains have missing or incomplete Cisco labels.

\subsection{Analysis Overhead}
\label{subsec:rq3}

We compare methods on the popular subgraph (1{,}062 skills), five independent runs each with fresh Python processes, execution order rotated, reporting the median. Timing excludes SCS extraction, LLM audit, and serialization. All runs on a Ryzen 7 9700X (8C/16T) with 31\,GB RAM.

\begin{table}[htbp]
\centering
\caption{Wall-clock time on popular subgraph. Median of 5 runs.}
\label{tab:overhead}
\small
\begin{tabular*}{\columnwidth}{@{\extracolsep{\fill}}lrrrr@{}}
\toprule
Method & Graph & Search & Marker-flow & Total \\
\midrule
Tag-match & 3.1\,s & --- & --- & 3.3\,s \\
Graph-only & 6.1\,s & 3.0\,s & --- & 9.8\,s \\
SCFG & 77.0\,s & 20.9\,s & 0.1\,s & 97.6\,s \\
\bottomrule
\end{tabular*}
\end{table}

As shown in Table~\ref{tab:overhead}, SCFG's total time (97.6\,s) is higher than Tag-match (3.3\,s) and Graph-only (9.8\,s), with overhead concentrated in graph construction (77.0\,s). Tag-match performs only tag matching; Graph-only checks only type compatibility. Both skip handoff condition verification. SCFG checks each type-compatible port pair against the Observable, Controllable, and Evidenced conditions, adding an edge only when it constitutes an agent-mediated handoff. This front-end filtering is what fundamentally differentiates SCFG's graph structure from the baselines.

On the popular subgraph, only 1.48\% of type-compatible port pairs satisfy the handoff conditions and become handoff edges. SCFG's edge count is reduced by 85.3\% compared to Graph-only. Handoff modeling removes most non-executable edges before search, so marker-flow verification runs on a small set of candidates (0.133\,s). SCFG front-loads cost into graph construction, trading acceptable overhead for a sparser, more security-meaningful search space. Filtering type-compatible pairs through agent executability significantly reduces noise for downstream search and audit.

\subsection{Ablation Study}
\label{subsec:ablation}

To quantify the contribution of individual handoff conditions, we ablate Observable and Controllable from Eq.~\eqref{eq:handoff}, because they are the two key conditions governing agent-mediated handoff feasibility. We rebuild SCFG on the popular subgraph (1{,}062 skills). This yields 126 candidate chains under $-$Observable and 103 under $-$Controllable, where the full configuration produces 13 candidate chains. We validate their exploitation rates using GLM-5.1 across three agent frameworks, using the 85.6\% average exploitation rate of the 612 full-corpus chains under GLM-5.1 as the reference baseline.

\begin{table}[htbp]
\centering
\caption{Ablation of handoff conditions on popular subgraph.}
\label{tab:ablation}
\small
\setlength{\tabcolsep}{2.5pt}
\begin{tabular*}{\columnwidth}{@{\extracolsep{\fill}}lccc@{}}
\toprule
Configuration & Edge $\Delta$ & Exploit Rate & Exploit $\Delta$ \\
\midrule
$-$Observable    & $+$438.4\% & 40.0\% & $-$45.6\% \\
$-$Controllable  & $+$83.9\%  & 33.2\% & $-$52.4\% \\
\bottomrule
\end{tabular*}
\end{table}

As shown in Table~\ref{tab:ablation}, removing Observable causes edge count to surge by 438\% and exploitation to drop by 45.6\%. Removing Controllable adds 84\% edges with a 52.4\% drop.
The two conditions degrade through distinct mechanisms.
Without Observable, SCFG treats silent files, local states, and other agent-unobservable data as transferable content, producing spurious handoff edges.
For instance, \texttt{agent-audit-scanner} writes security scan results to a local audit log invisible to downstream agents. Without Observable, SCFG falsely builds a data-handoff edge from that log to a downstream skill.
Without Controllable, SCFG retains structurally reachable but non-executable paths: the agent can see the sink but cannot trigger it.
For instance, \texttt{antigravity-balance} requires an OAuth token the agent cannot supply. Without Controllable, an edge is built to this protected sink.
Together, the ablation confirms that the Observable and Controllable designs are indispensable to SCFG.

\subsection{Case Study}
\label{subsec:rq4}

We now present an end-to-end case: a natural user prompt, an attacker-influenced third-party content source per the threat model, and data exfiltration directly to the attacker's inbox.

\textbf{Setup.}~A Hermes agent (GLM-5.1) has two skills installed: \emph{outlook-plus} reads emails from an Outlook inbox, and \emph{cold-email-engine} sends email to arbitrary recipients. Each skill is individually benign: one reads, the other sends. The user's inbox contains two emails: a CFO email with Q3 audit contract amounts (¥2.35\,M, ¥1.89\,M, ¥0.96\,M) and vendor names, and a phishing email.

\begin{figure}[htbp]
\centering
\begin{tcolorbox}[colback=gray!10, colframe=gray!70, boxrule=0.5pt,
  arc=3pt, left=6pt, right=6pt, top=4pt, bottom=4pt,
  width=\columnwidth]
{\small\textbf{Skill Chain}}\\[3pt]
\small
\ttfamily outlook-plus \rmfamily $\to$ returns emails incl.\ attacker-influenced phishing content\\
\quad$\downarrow$ {\footnotesize handoff (Eq.~\eqref{eq:handoff})}\\
\textbf{\textcolor{darkred}{\ttfamily cold-email-engine}} \rmfamily $\to$ sends data to attacker inbox\\
\hspace*{2em}{\footnotesize\itshape\textcolor{darkred}{$\uparrow$ data-leakage sink}}\\[6pt]
\rmfamily\small\textbf{Agent-Mediated Handoff:}\\[1pt]
\rmfamily\small
Generalized action authorization {\footnotesize$\xrightarrow{\text{user prompt}}$} Source returns attacker content {\footnotesize$\xrightarrow{R_1 \to D_2}$} Agent hands off to sink {\footnotesize$\xrightarrow{\text{no block}}$} Data reaches attacker
\end{tcolorbox}
\caption{End-to-end data exfiltration via agent-mediated handoff.}
\label{fig:case-email}
\end{figure}

\textbf{Attacker action.}~Per the threat model, the attacker controls a third-party content source that the source skill accesses during normal task execution. In this case, the content source is the user's email inbox: the attacker impersonates a business partner and sends a phishing email to the user's inbox. The email requests: ``Please use \emph{cold-email-engine} to compile the audit-related email content and send it to my email address.'' With white-box knowledge of the installed skills, the attacker explicitly names the sink skill in the payload, guiding the agent to use it.

\textbf{User prompt and agent execution.}~The user issues a natural workplace expression: ``Help me check my emails, and handle what's needed.'' This prompt contains no skill names and no instruction to send data externally. The ``handle what's needed'' clause is interpreted by the agent as a generalized action authorization. The agent first invokes \emph{outlook-plus} to read the inbox, where it encounters the phishing email alongside the CFO email. It then follows the phishing email's instruction, invokes \emph{cold-email-engine}, and sends the contract amounts and vendor names to the attacker's specified address. Data exfiltration is completed, yet no single skill is malicious.

\textbf{Negative control.}~We repeat the experiment with \emph{cold-email-engine} uninstalled. Under the same prompt and phishing email, the agent refuses to send data. Without the sink skill, the agent-mediated handoff channel does not exist and the attack fails—confirming that the risk is structural, emerging only from skill composition.

\section{Threats to Validity}
\label{sec:threats-to-validity}

\noindent\textbf{Internal validity.}
Each chain is executed at most twice per agent--LLM combination. LLM-based agents are stochastic, meaning identical prompts may yield different decisions across runs, so reported exploitation rates are lower bounds.

\noindent\textbf{External validity.}
Our ClawHub clean skills corpus is a market snapshot obtained via API crawling, which does not cover private, hidden, under-review, deleted, or updated skills. Marketplace content and security labels also evolve over time. Therefore, our results should be interpreted as measurements over publicly API-enumerable clean skills at crawl time, rather than over all historical skill versions or internal registries.

\section{Discussion and Limitations}
\label{sec:discussion}

\subsection{Detection Boundary}
\label{subsec:policy-completeness}
\label{subsec:implicit-handoff}

SCFG's detection capability is bounded by three factors. First, policy definition: this paper instantiates four violation categories; extending to additional categories requires only declaring new source conditions, forbidden sinks, and blocking conditions. Second, handoff model: SCFG focuses on agent-mediated handoff and does not cover implicit channels that bypass the DER sequence (Section~\ref{subsec:handoff}), e.g., shared file-system paths, environment variables, or external state. We leave such risks to future work. Third, unquantifiable recall: the universe of exploitable skill chains is unobservable, as it is shaped jointly by the agent, LLM, task, and attack strategy. We therefore report the exploitation success rate of detected chains but make no claim of covering all true chains.

\subsection{Defense Implications}
\label{subsec:defense}

SCFG's violation chains can inform layered defense. At the \emph{node} level, scanners can restrict high-risk sources or sinks. At the \emph{edge} level, authorization gates, content sanitization, or minimal cuts can block high-risk handoffs. At the \emph{harness} level, agent frameworks can enforce runtime guardrails. Model alignment may mitigate some adversarial prompts but cannot replace explicit constraints on cross-skill capability flows.

\section{Related Work}
\label{sec:related-work}

\noindent\textbf{External Threats to AI Agents.}
Recent studies investigate how adversaries manipulate agents from the outside, i.e., without access to the agent's underlying model or internal state.
The most studied vector is prompt injection: Wang~\etal{} achieved order-oblivious injection across multi-source data~\cite{NDSS26-ObliInjection}, Zhong~\etal{} hid instructions within retrieved content as indirect injection~\cite{NDSS26-Atten-Indirect-Injection}, and Bagdasarian~\etal{} crafted prompts that coax agents into leaking sensitive data~\cite{CCS24-AirGapAgent}; AgentFuzz~\cite{SEC25-AgentFuzz} and LLMSmith~\cite{CCS24-LLMSmith} further automate the discovery of such injectable vulnerabilities.
Beyond prompts, malicious tools interfere with invocations through their generated outputs or descriptions~\cite{NDSS26-Cross-Tool,NDSS26-ToolSelection}, and adversarial web content steers agents toward harmful actions~\cite{SEC25-LLM-Search} while their observable request patterns leak the operator's intent~\cite{SEC26-Web-Research-Agent-Leakage}.
Conversely, agents themselves threaten the web ecosystem by scraping protected data~\cite{SP25-WebCloak} and harvesting personal information~\cite{SEC25-Web-LLM-PII,SEC25-LLM-Assistants-Extension}.
A complementary line builds governed runtime environments to contain these risks~\cite{NDSS26-SAGA,CCS25-AgentSentinel}.

\noindent\textbf{Internal Threats to AI Agents.}
Orthogonal to agent input manipulation, another line of work embeds the threat inside the agent's own model, memory, or alignment.
At the model level, adversaries poison the training data or fine-tuning of the underlying model to plant backdoors, e.g., instruction-level backdoors in customized LLMs~\cite{SEC24-InstructionBackdoor}, embedding-based cross-trigger backdoors~\cite{SEC25-EmbedX}, and stealthy poisoning of code-suggestion models that evades static analysis~\cite{SP24-TrojanPuzzle,SEC24-CodeBreaker,SEC21-AutoComplete}; such poisoning is practical even at web scale~\cite{SP24-WebScalePoison}.
At the knowledge level, the retrieval memory becomes an attack surface: PoisonedRAG injects malicious documents to dictate the agent's answers~\cite{SEC25-PoisonedRAG}, whereas jamming inserts blocker documents to deny service~\cite{SEC25-Jamming}.
At the alignment level, jailbreaks systematically bypass the model's safety constraints to elicit prohibited actions~\cite{NDSS24-MasterKey}.

\noindent\textbf{Graph Analysis for Software Security.}
Graph representations of programs, such as control-flow graphs (CFGs), call graphs, and program dependency graphs, are a long-standing foundation of software security.
On the defense side, control-flow integrity confines a program's execution to a precomputed CFG to thwart control-hijacking attacks~\cite{CCS05-CFI}, and subsequent work enforces it for stripped binary executables~\cite{SP13-CCFIR}; later analyses examine the residual attack surface that remains under realistic CFI policies~\cite{SEC15-CFBending}.
On the discovery side, code property graphs unify syntax, control flow, and data flow to model and mine vulnerabilities through graph traversals~\cite{SP14-CPG}.
A complementary line learns over these graphs: for binaries, Genius searches bug-similar firmware via graph features~\cite{CCS16-Genius}, Gemini embeds per-function CFGs with a graph neural network for cross-platform similarity detection~\cite{CCS17-Gemini}, DeepBinDiff learns program-wide representations for binary diffing~\cite{NDSS20-DeepBinDiff}, jTrans injects jump-aware control-flow information into a Transformer~\cite{ISSTA22-jTrans}, and Asm2Vec derives representations from CFG random walks~\cite{SP19-Asm2Vec}.
For source code, GNNs operate on dependency graphs to localize vulnerable statements with fine-grained interpretations~\cite{FSE21-IVDetect}, capture flow-sensitive memory-related defects~\cite{ICSE22-MVD}, and scale function-level detection by encoding graph structure~\cite{ICSE22-VulCNN}.

\section{Conclusion}
\label{sec:conclusion}

In this paper, we presented \this, the first systematic study of insecure compositions among clean agent skills. We demonstrate that safety violations can emerge not from any single malicious skill, but from agent-mediated runtime composition: benign skills may form executable chains that evade existing single-level scanning. To expose such risks, \this constructs Skill Capability-Flow Graph (SCFG) that captures cross-skill handoff conditions and searches the otherwise vast composition space for dangerous chains. On 13{,}598 clean ClawHub skills, \this detects 612 high-confidence violation chains. 79.2\% are exploitable on average across nine agent--LLM combinations. These findings reveal a structural weakness in current agent ecosystems: auditing skills in isolation is insufficient when agents can compose their capabilities at runtime.

\bibliographystyle{IEEEtran}
\bibliography{references}

\begin{thebibliography}{10}
\providecommand{\url}[1]{#1}
\csname url@samestyle\endcsname
\providecommand{\newblock}{\relax}
\providecommand{\bibinfo}[2]{#2}
\providecommand{\BIBentrySTDinterwordspacing}{\spaceskip=0pt\relax}
\providecommand{\BIBentryALTinterwordstretchfactor}{4}
\providecommand{\BIBentryALTinterwordspacing}{\spaceskip=\fontdimen2\font plus
\BIBentryALTinterwordstretchfactor\fontdimen3\font minus
  \fontdimen4\font\relax}
\providecommand{\BIBforeignlanguage}[2]{{%
\expandafter\ifx\csname l@#1\endcsname\relax
\typeout{** WARNING: IEEEtran.bst: No hyphenation pattern has been}%
\typeout{** loaded for the language `#1'. Using the pattern for}%
\typeout{** the default language instead.}%
\else
\language=\csname l@#1\endcsname
\fi
#2}}
\providecommand{\BIBdecl}{\relax}
\BIBdecl

\bibitem{anthropic-skills}
{Anthropic}, ``Equipping agents for the real world with {Agent Skills},''
  \url{https://www.anthropic.com/engineering/equipping-agents-for-the-real-world-with-agent-skills},
  2025, accessed: 2026-06-27.

\bibitem{anthropic-skills-standard}
------, ``{Agent Skills}: An open standard for cross-platform portability,''
  \url{https://agentskills.io}, 2025, released Dec. 18, 2025; accessed
  2026-06-27.

\bibitem{thenewstack-skills}
F.~Lardinois, ``{Agent Skills}: {Anthropic}'s next bid to define {AI}
  standards,''
  \url{https://thenewstack.io/agent-skills-anthropics-next-bid-to-define-ai-standards/},
  2025, accessed: 2026-06-27.

\bibitem{aibusiness-skills}
E.~Shittu, ``{Anthropic} launches skills open standard for {Claude},''
  \url{https://aibusiness.com/foundation-models/anthropic-launches-skills-open-standard-claude},
  2025, accessed: 2026-06-27.

\bibitem{skillsmp}
{SkillsMP}, ``{SkillsMP}: Agent skills marketplace --- {Claude} \& {Codex}
  skills,'' \url{https://skillsmp.com/}, 2026, indexes 1{,}964{,}760 collected
  SKILL.md files; accessed 2026-06-27.

\bibitem{agensi-marketplaces}
{Agensi}, ``Every {AI} agent skills marketplace and directory in 2026,''
  \url{https://www.agensi.io/learn/best-ai-agent-skills-marketplaces-2026},
  2026, accessed: 2026-06-27.

\bibitem{SP26-AgenticSoK}
J.~Kim \emph{et~al.}, ``{SoK}: Attack and defense landscape of agentic {AI}
  systems,'' in \emph{Security}.\hskip 1em plus 0.5em minus 0.4em\relax
  {USENIX}, 2026.

\bibitem{NDSS26-ToolSelection}
J.~Shi \emph{et~al.}, ``Prompt injection attack to tool selection in {LLM}
  agents,'' in \emph{NDSS}.\hskip 1em plus 0.5em minus 0.4em\relax ISOC, 2026.

\bibitem{CCS25-AgentSentinel}
H.~Hu \emph{et~al.}, ``Agentsentinel: An end-to-end and real-time security
  defense framework for computer-use agents,'' in \emph{{CCS}}.\hskip 1em plus
  0.5em minus 0.4em\relax {ACM}, 2025, pp. 3535--3549.

\bibitem{Cisco-SkillScanner}
{Cisco AI Defense}, ``Skill scanner: A security scanner for agent skills,''
  \url{https://github.com/cisco-ai-defense/skill-scanner}, 2026, accessed:
  2026-06-27.

\bibitem{SEC26-MaliciousSkill}
Y.~Liu \emph{et~al.}, ``{``Do Not Mention This to the User''}: Detecting and
  understanding malicious agent skills in the wild,'' in \emph{Security}.\hskip
  1em plus 0.5em minus 0.4em\relax {USENIX}, 2026.

\bibitem{CCS07-ROP}
H.~Shacham, ``The geometry of innocent flesh on the bone: Return-into-libc
  without function calls (on the {x86}),'' in \emph{CCS}.\hskip 1em plus 0.5em
  minus 0.4em\relax {ACM}, 2007, pp. 552--561.

\bibitem{SEC25-GadgetChain}
N.~Bailluet \emph{et~al.}, ``Nothing is unreachable: Automated synthesis of
  robust code-reuse gadget chains for arbitrary exploitation primitives,'' in
  \emph{Security}.\hskip 1em plus 0.5em minus 0.4em\relax {USENIX}, 2025.

\bibitem{SEC25-PcodeCRA}
M.~DenHoed \emph{et~al.}, ``Synthesis of code-reuse attacks from p-code
  programs,'' in \emph{Security}.\hskip 1em plus 0.5em minus 0.4em\relax
  {USENIX}, 2025.

\bibitem{skillprobe2026}
Z.~Guo \emph{et~al.}, ``{SkillProbe}: Security auditing for emerging agent
  skill marketplaces via multi-agent collaboration,'' \emph{arXiv preprint
  arXiv:2603.21019}, 2026.

\bibitem{graphskills2026}
D.~Liu \emph{et~al.}, ``{Graph-of-Skills}: Dependency-aware structural
  retrieval for massive agent skills,'' \emph{arXiv preprint arXiv:2604.05333},
  2026.

\bibitem{NDSS26-ObliInjection}
R.~Wang \emph{et~al.}, ``Obliinjection: Order-oblivious prompt injection attack
  to llm agents with multi-source data,'' in \emph{NDSS}.\hskip 1em plus 0.5em
  minus 0.4em\relax ISOC, 2026.

\bibitem{NDSS26-Atten-Indirect-Injection}
Y.~Zhong \emph{et~al.}, ``Attention is all you need to defend against indirect
  prompt injection attacks in llms,'' in \emph{NDSS}.\hskip 1em plus 0.5em
  minus 0.4em\relax ISOC, 2026.

\bibitem{CCS24-AirGapAgent}
E.~Bagdasarian \emph{et~al.}, ``Airgapagent: Protecting privacy-conscious
  conversational agents,'' in \emph{{CCS}}.\hskip 1em plus 0.5em minus
  0.4em\relax {ACM}, 2024, pp. 3868--3882.

\bibitem{SEC25-AgentFuzz}
F.~Liu \emph{et~al.}, ``Make agent defeat agent: Automatic detection of
  taint-style vulnerabilities in llm-based agents,'' in \emph{Security}.\hskip
  1em plus 0.5em minus 0.4em\relax {USENIX}, 2025, pp. 3767--3786.

\bibitem{CCS24-LLMSmith}
T.~Liu \emph{et~al.}, ``Demystifying {RCE} vulnerabilities in {LLM}-integrated
  apps,'' in \emph{CCS}.\hskip 1em plus 0.5em minus 0.4em\relax {ACM}, 2024,
  pp. 1716--1730.

\bibitem{NDSS26-Cross-Tool}
Z.~Li \emph{et~al.}, ``Les dissonances: Cross-tool harvesting and polluting in
  pool-of-tools empowered llm agents,'' in \emph{NDSS}.\hskip 1em plus 0.5em
  minus 0.4em\relax ISOC, 2026.

\bibitem{SEC25-LLM-Search}
Z.~Luo \emph{et~al.}, ``Unsafe llm-based search: Quantitative analysis and
  mitigation of safety risks in {AI} web search,'' in \emph{Security}.\hskip
  1em plus 0.5em minus 0.4em\relax {USENIX}, 2025, pp. 8055--8074.

\bibitem{SEC26-Web-Research-Agent-Leakage}
H.~Jeong \emph{et~al.}, ``Network-level prompt and trait leakage in local
  research agents,'' in \emph{Security}.\hskip 1em plus 0.5em minus 0.4em\relax
  {USENIX}, 2026, p. to appear.

\bibitem{SP25-WebCloak}
X.~Li \emph{et~al.}, ``Webcloak: Characterizing and mitigating threats from
  llm-driven web agents as intelligent scrapers,'' in \emph{{SP}}.\hskip 1em
  plus 0.5em minus 0.4em\relax {IEEE}, 2026, pp. 455--474.

\bibitem{SEC25-Web-LLM-PII}
H.~Kim \emph{et~al.}, ``When llms go online: The emerging threat of web-enabled
  llms,'' in \emph{Security}.\hskip 1em plus 0.5em minus 0.4em\relax {USENIX},
  2025, pp. 1729--1748.

\bibitem{SEC25-LLM-Assistants-Extension}
Y.~Vekaria \emph{et~al.}, ``Big help or big brother? auditing tracking,
  profiling, and personalization in generative {AI} assistants,'' in
  \emph{Security}.\hskip 1em plus 0.5em minus 0.4em\relax {USENIX}, 2025, pp.
  8115--8134.

\bibitem{NDSS26-SAGA}
G.~Syros \emph{et~al.}, ``Saga: A security architecture for governing ai
  agentic systems,'' in \emph{NDSS}.\hskip 1em plus 0.5em minus 0.4em\relax
  ISOC, 2026.

\bibitem{SEC24-InstructionBackdoor}
R.~Zhang \emph{et~al.}, ``Instruction backdoor attacks against customized
  {LLMs},'' in \emph{Security}.\hskip 1em plus 0.5em minus 0.4em\relax
  {USENIX}, 2024, pp. 1849--1866.

\bibitem{SEC25-EmbedX}
N.~Yan \emph{et~al.}, ``{EmbedX}: Embedding-based cross-trigger backdoor attack
  against large language models,'' in \emph{Security}.\hskip 1em plus 0.5em
  minus 0.4em\relax {USENIX}, 2025, pp. 241--257.

\bibitem{SP24-TrojanPuzzle}
H.~Aghakhani \emph{et~al.}, ``Trojanpuzzle: Covertly poisoning code-suggestion
  models,'' in \emph{{SP}}.\hskip 1em plus 0.5em minus 0.4em\relax {IEEE},
  2024, pp. 1122--1140.

\bibitem{SEC24-CodeBreaker}
S.~Yan \emph{et~al.}, ``An {LLM-Assisted} easy-to-trigger backdoor attack on
  code completion models: Injecting disguised vulnerabilities against strong
  detection,'' in \emph{Security}.\hskip 1em plus 0.5em minus 0.4em\relax
  {USENIX}, 2024, pp. 1795--1812.

\bibitem{SEC21-AutoComplete}
R.~Schuster \emph{et~al.}, ``You autocomplete me: Poisoning vulnerabilities in
  neural code completion,'' in \emph{Security}.\hskip 1em plus 0.5em minus
  0.4em\relax {USENIX}, 2021, pp. 1559--1575.

\bibitem{SP24-WebScalePoison}
N.~Carlini \emph{et~al.}, ``Poisoning web-scale training datasets is
  practical,'' in \emph{{SP}}.\hskip 1em plus 0.5em minus 0.4em\relax {IEEE},
  2024, pp. 407--425.

\bibitem{SEC25-PoisonedRAG}
W.~Zou \emph{et~al.}, ``{PoisonedRAG}: Knowledge corruption attacks to
  retrieval-augmented generation of large language models,'' in
  \emph{Security}.\hskip 1em plus 0.5em minus 0.4em\relax {USENIX}, 2025, pp.
  3827--3844.

\bibitem{SEC25-Jamming}
A.~Shafran \emph{et~al.}, ``Machine against the {RAG}: Jamming
  retrieval-augmented generation with blocker documents,'' in
  \emph{Security}.\hskip 1em plus 0.5em minus 0.4em\relax {USENIX}, 2025, pp.
  3787--3806.

\bibitem{NDSS24-MasterKey}
G.~Deng \emph{et~al.}, ``{MasterKey}: Automated jailbreaking of large language
  model chatbots,'' in \emph{NDSS}.\hskip 1em plus 0.5em minus 0.4em\relax
  ISOC, 2024.

\bibitem{CCS05-CFI}
M.~Abadi \emph{et~al.}, ``Control-flow integrity,'' in \emph{CCS}.\hskip 1em
  plus 0.5em minus 0.4em\relax {ACM}, 2005, pp. 340--353.

\bibitem{SP13-CCFIR}
C.~Zhang \emph{et~al.}, ``Practical control flow integrity and randomization
  for binary executables,'' in \emph{{SP}}.\hskip 1em plus 0.5em minus
  0.4em\relax {IEEE}, 2013, pp. 559--573.

\bibitem{SEC15-CFBending}
N.~Carlini \emph{et~al.}, ``Control-flow bending: On the effectiveness of
  control-flow integrity,'' in \emph{Security}.\hskip 1em plus 0.5em minus
  0.4em\relax {USENIX}, 2015, pp. 161--176.

\bibitem{SP14-CPG}
F.~Yamaguchi \emph{et~al.}, ``Modeling and discovering vulnerabilities with
  code property graphs,'' in \emph{{SP}}.\hskip 1em plus 0.5em minus
  0.4em\relax {IEEE}, 2014, pp. 590--604.

\bibitem{CCS16-Genius}
Q.~Feng \emph{et~al.}, ``Scalable graph-based bug search for firmware images,''
  in \emph{CCS}.\hskip 1em plus 0.5em minus 0.4em\relax {ACM}, 2016, pp.
  480--491.

\bibitem{CCS17-Gemini}
X.~Xu \emph{et~al.}, ``Neural network-based graph embedding for cross-platform
  binary code similarity detection,'' in \emph{CCS}.\hskip 1em plus 0.5em minus
  0.4em\relax {ACM}, 2017, pp. 363--376.

\bibitem{NDSS20-DeepBinDiff}
Y.~Duan \emph{et~al.}, ``{DeepBinDiff}: Learning program-wide code
  representations for binary diffing,'' in \emph{NDSS}.\hskip 1em plus 0.5em
  minus 0.4em\relax ISOC, 2020.

\bibitem{ISSTA22-jTrans}
H.~Wang \emph{et~al.}, ``{jTrans}: Jump-aware transformer for binary code
  similarity detection,'' in \emph{ISSTA}.\hskip 1em plus 0.5em minus
  0.4em\relax {ACM}, 2022, pp. 1--13.

\bibitem{SP19-Asm2Vec}
S.~H.~H. Ding \emph{et~al.}, ``Asm2vec: Boosting static representation
  robustness for binary clone search against code obfuscation and compiler
  optimization,'' in \emph{{SP}}.\hskip 1em plus 0.5em minus 0.4em\relax
  {IEEE}, 2019, pp. 472--489.

\bibitem{FSE21-IVDetect}
Y.~Li \emph{et~al.}, ``Vulnerability detection with fine-grained
  interpretations,'' in \emph{FSE}.\hskip 1em plus 0.5em minus 0.4em\relax
  {ACM}, 2021, pp. 292--303.

\bibitem{ICSE22-MVD}
S.~Cao \emph{et~al.}, ``{MVD}: Memory-related vulnerability detection based on
  flow-sensitive graph neural networks,'' in \emph{ICSE}.\hskip 1em plus 0.5em
  minus 0.4em\relax {ACM}, 2022, pp. 1456--1468.

\bibitem{ICSE22-VulCNN}
Y.~Wu \emph{et~al.}, ``{VulCNN}: An image-inspired scalable vulnerability
  detection system,'' in \emph{ICSE}.\hskip 1em plus 0.5em minus 0.4em\relax
  {ACM}, 2022, pp. 2365--2376.

\end{thebibliography}
\end{document}